\documentclass[letterpaper, 10 pt, conference]{ieeeconf}  
\IEEEoverridecommandlockouts
\usepackage{cite}
\usepackage{amsmath,amssymb,amsfonts}
\usepackage{graphicx}
\usepackage{textcomp}
\usepackage{xcolor}
\usepackage{accents}
\usepackage[mathscr]{euscript}
\usepackage{mathrsfs}
\usepackage{mathtools}
\usepackage{textcomp, gensymb}
\usepackage{booktabs}
\usepackage{multirow}
\usepackage{array}
\usepackage{algorithm, algpseudocode}
\usepackage[capitalize]{cleveref}

\begin{document}

\title{Using Flexibility Envelopes for the Demand-Side Hierarchical Optimization of District Heating Networks}

\author{Audrey~Blizard,~Colin~N.~Jones,~and~Stephanie~Stockar%
\thanks{This material is based upon work supported by the National Science Foundation Graduate Research Fellowship Program under Grant No. DGE-1343012}%
\thanks{Audrey Blizard (blizard.1@osu.edu) and Stephanie Stockar (stockar.1@osu.edu) are with the  Department of Mechanical and Aerospace Engineering and the Center for Automotive Research, The Ohio State University, 930 Kinnear Road, Columbus, OH 43212 USA}%
\thanks{Colin N. Jones (colin.jones@epfl.ch) is with LA, EPFL, 1015 Lausanne, Switzerland}}

\maketitle

\begin{abstract}
The demand-side control of district heating networks is notoriously challenging due to the large number of connected users and the high number of states to be considered. To overcome these challenges, this paper presents a hierarchical optimization scheme using the flexibility in heating demand provided by the users to improve the performance of the network. This hierarchical scheme relies on a low level controller to calculate the costs for a subsystem over a given set of potential pressure drops for that subsystem. The high level controller then uses these calculated costs to determine the optimal set of pressure drops for every subgraph of the partitioned network. The proposed hierarchical optimization scheme is demonstrated on a representative 20 user district heating network, resulting in a 67\% reduction in bypass mass flow while ensuring all network users stay within 2 \degree C of their desired nominal temperatures.
\end{abstract}

\section{Introduction}
In 2022, nearly half of the energy used in buildings was for space and water heating \cite{BuildingsEnergySystem}. District heating networks (DHNs) are a promising alternative to traditional methods of supplying this heat. By utilizing economy of scale and enabling the easier integration of renewable energy sources, DHNs are a valuable tool to reduce emissions and lower costs, especially in cities. However, existing control methods are not sufficient to capitalize on the full energy saving potential of these networks and advanced controllers are needed to further improve their performance. \par
Because DHNs are networked systems with many connected users, they provide more flexibility in how heat can be delivered as compared to single-user heating systems, while still ensuring the comfort of the connected users. This flexibility has been exploited by advanced controllers in a variety of ways \cite{vandermeulenControllingDistrictHeating2018}. For example, it has been used to better meet the electricity demands of the transmission system in a DHN fed by a combined heat and power system \cite{delorenziPredictiveControlCombined2022}. Additionally, the flexibility has been used to reduce total operating costs by considering variable pricing in the electricity supply \cite{vanhoudtActiveControlStrategy2018} and incorporate additional renewable energy sources \cite{jiangExploitingFlexibilityDistrict2020}. \par
However, the focus of research on the control of DHNs has been on the interconnection between the network and the heat source to either reduce energy consumption, shift heat demand, or decrease operating costs, by aggregating the network into a single flexible consumer. Significantly less research has been conducted on demand-side management, which aims to modulate the individual user's behaviors to reduce the total energy consumption of the network. Initial studies have shown that demand-side management is a promising method to reduce energy consumption, with one study showing a 35\% reduction in energy consumption is possible \cite{delorenziSetupTestingSmart2020}. This work proposes combining the flexibility provided by the buildings connected to the network with a demand-side control approach.\par
Centralized control of the individual users in DHNs is challenging, especially as the system scales due the large number of users and geographic dispersion \cite{vandermeulenControllingDistrictHeating2018}. Therefore, instead of relying on a centralized controller, a novel hierarchical control structure is proposed. In this structure, the network is decomposed into manageable subsystems to be optimized with assumptions introduced on the total pressure losses in the subsystems. \par
The low-level controller utilizes the flexibility of the local buildings to minimize the control objective for the local subsystem. Here, the available flexibility of a building will be quantified through the use of flexibility envelopes \cite{reyndersEnergyFlexibleBuildings2018}, which provide a measure of a building's ability to temporarily accept a higher or lower heat supply while still remaining in an acceptable temperature range. The concept of flexibility envelopes has been proven successful for the integration of renewable energy sources \cite{liImprovingOperationalFlexibility2020} and in the control of buildings demand for peak shaving operations \cite{gasserPredictiveEnergyManagement2021}.
The high-level controller then uses the subsystems' costs to determine the optimal set of pressure losses and mass flow rates for every subsystem. The resulting hierarchical structure is demonstrated on a 20 user DHN, with the objective to minimize bypass flow rate as compared to the nominal case.   
\section{System Model}
\begin{figure}
    \centering
    \includegraphics[width = 1\columnwidth]{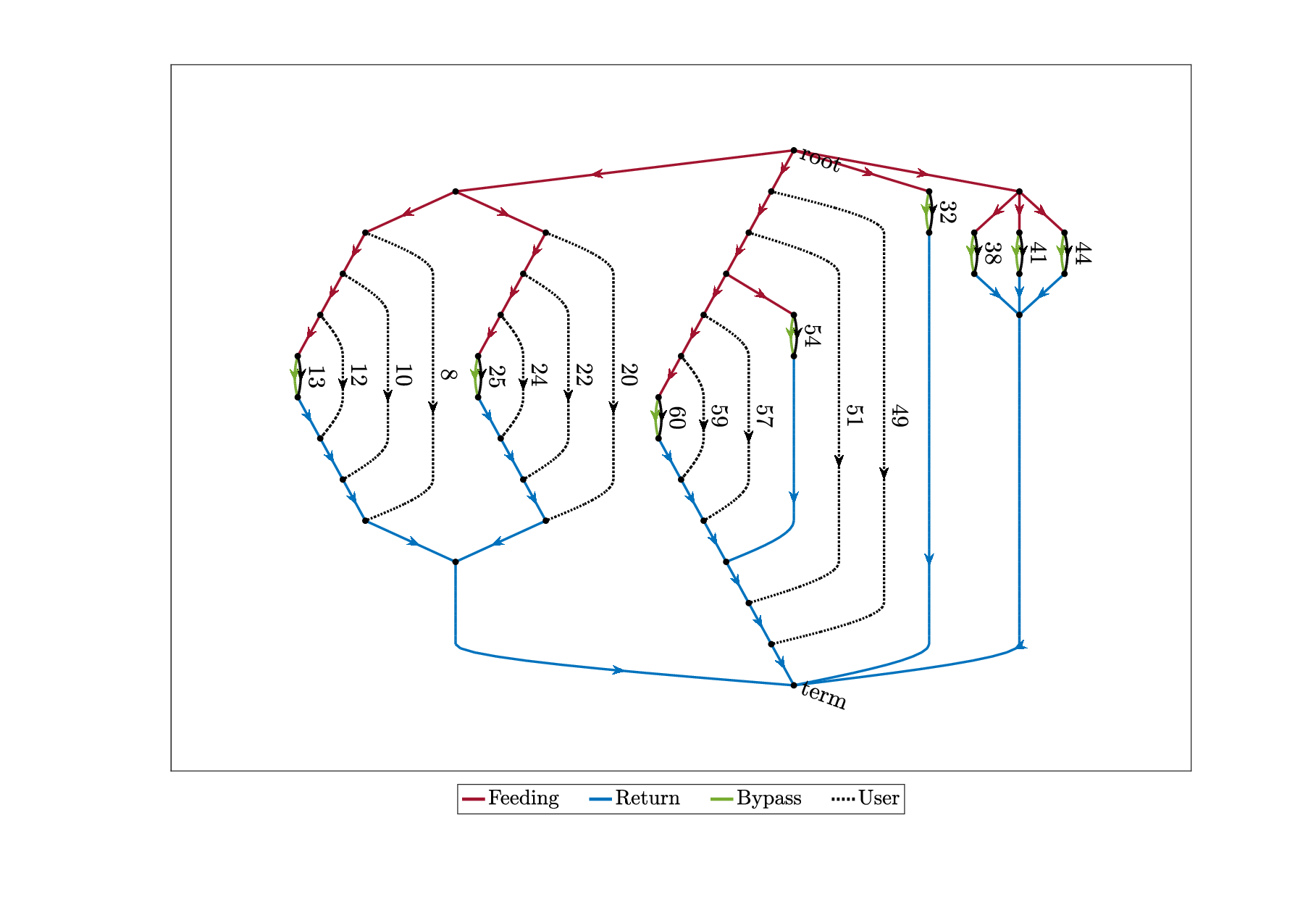}
    \caption{Sample graph for an 18-user DHN.}
    \label{fig:graph}
\end{figure}

The three main elements of DHNs that must be modeled to accurately describe the thermal behavior of the network are 1) the heating plant, which supplies water with a certain mass flow rate and temperature, 2) the feeding/return network, which delivers the heated water to the users and returns the cooled water to the plant, and 3) the users, which extract heat from the network to meet their heat demands. It is convenient to represent a DHN as a directed graph $\mathcal{G} = (\mathcal{V},\mathcal{E})$, where $\mathcal{E}$ is the set of edges, which can be decomposed into four disjoint subsets, $\mathcal{E}=\{\mathcal{E}_F,\mathcal{E}_R,\mathcal{E}_{By},\mathcal{E}_{U}\}$ for the feeding lines, return lines, bypass segments, and user segments, respectively, and $\mathcal{V}$ is the set of nodes connecting the pipes in the network. Associated with the graph $\mathcal{G}$ are an adjacency matrix, $\Gamma$, and incidence matrix, $\Lambda$, which are both methods of representing the connections between the nodes and edges in a graph. In the graph, the plant is modeled by a pair of nodes $v_{root}, v_{term}$ where $\text{indegree}(v_{root})=0$ and $\text{outdegree}(v_{term})=0$. A sample graph of a DHN is shown in \cref{fig:graph}, with the sets of edges highlighted. The behavior of the network components is modeled via a first principle approach as follows.
\subsection{Thermal Model}
A reduced version of the graph-based model, originally developed by the authors is used to model the thermodynamic response of the network pipes \cite{blizardGraphBasedTechniqueAutomated2024}. The temperature of an individual pipe segment is found using a well-mixed approach according to
\begin{equation}
    \label{eq:Tpipec}
    \frac{d}{dt}T=c_1\cdot\dot{m}\cdot T_{in} + c_2 T_{amb} - (c_1\cdot\dot{m}+c_2) T
\end{equation}
where $\dot{m}$ is the mass flow rate through the pipe, $T_{in}$ is the inlet temperature in the pipe segment, $T_{amb}$ is the ambient temperature, $T$ is the temperature in the pipe segment, and
\begin{subequations}
\label{eq:c}
\begin{equation}
    {c_1} = \frac{1}{\rho V}
\end{equation}
\begin{equation}
    {c_2} = \frac{hA_s}{\rho c_pV}
\end{equation}
\end{subequations}
where $\rho$ and $c_p$ are the density and constant pressure heat capacity of the operating fluid respectively, and $V$ and $hA_s$ are the volume and the heat transfer coefficient of the pipe.\par
Additionally, it is assumed the temperature of the user edge is a known constant, $T_{setR}$. This assumption is valid as this temperature can be directly modulated via the speed the operating fluid is re-circulated through the heat exchanger, and the dynamics of the user edges are much faster than those of the rest of the network pipes, allowing these edges to be neglected in the dynamic temperature model.\par
From this assumption and the previously developed model, the state space representation of the entire network takes the form
\begin{equation}
\label{eq:AugSS}
    \frac{d}{dt}T_{\mathcal{E}_N}  = A(\dot{m}_{\mathcal{E}_N})T_{\mathcal{E}_N} +B \begin{bmatrix}T_0\\T_{setR}\\T_{amb}\end{bmatrix}
\end{equation}
where $\mathcal{E}_N = \mathcal{E}\backslash\mathcal{E}_U$ is the set of non-user edges, $T_{\mathcal{E}_N}$ is the vector of temperatures in these pipes, $T_0$ is the supply temperature, and the $A$ and $B$ matrices can be found by placing the mass flow rates and $c_1,c_2,c_3$ coefficients based on the interconnections of the pipes in the network.
\subsection{Fluids Model}
The mass flow rate in the network can be fully resolved with three sets of equations: the pressure loss equations for each pipe segment, the network conservation of mass equation, and the network pressure balance equation \cite{ulanickiSimplificationWaterDistribution1996}. The pressure loss, $\Delta P$, in an individual pipe segment can be modeled by
\begin{equation}
\label{eq:dP}
    \Delta P = \zeta\dot{m}^2
\end{equation}
where $\zeta$ is the total pressure loss coefficient, determined based on the pipe's characteristics, such as length, diameter, and material.\par
The conservation of mass is enforced by
\begin{equation}
    \Lambda \dot{m}_{\mathcal{E}} = \dot{m}_{\mathcal{V}}
\end{equation}
where $\dot{m}_{\mathcal{E}}$ is the vector of mass flow rates in the edges and $\dot{m}_{\mathcal{V}}$ is the vector of the mass flow rates in the nodes, given by
\begin{equation}
    \dot{m}_{\mathcal{V}}(i) =\begin{cases}
    \dot{m}_0 & \text{if } v(i) = v_{root}\\
    -\dot{m}_0 & \text{if } v(i) = v_{term}\\
    0 &\text{otherwise}
    \end{cases}
\end{equation}
where $\dot{m}_0$ is the supply mass flow rate.\par
Finally, the pressure balance in the network can be found using 
\begin{equation}
    \Delta P_{\mathcal{E}} = \Lambda^T P_{\mathcal{V}}
\end{equation}
where $\Delta P_{\mathcal{E}}$ is a vector of pressure losses in the edges and $P_{\mathcal{V}}$ is a vector of the pressure at every node where $P_{\mathcal{V}}(v_{term})=0$ serves as the reference pressure. 
\subsection{Flexible Building Model}
The users in the network are modeled as flexible consumers of heat, using a similar method to the one developed in Reynders et al. \cite{reyndersGenericCharacterizationMethod2017}. The range of flexibility in the heat demand, called the flexibility envelope is given by
\begin{equation}
    C\Delta T_L\leq\int_{t_0}^{t_f} \Delta\dot{Q}_p(t)\ dt+\Delta\dot{Q}_p(t_0)\leq C\Delta T_U
\end{equation}
where $C$ is the heat capacity of the building, $\Delta T_L$ and $\Delta T_U$ are the acceptable upper and lower temperature deviation from the nominal desired temperature (${T_B}_{nom}$), $\Delta\dot{Q}_p$ is the deviation from the nominal heat demand $\dot{Q}_{out}$, needed to keep the building at its desired temperature, given by
\begin{equation}
    \Delta\dot{Q}_p=\dot{Q}_p-\dot{Q}_{out}
\end{equation}
and $\Delta\dot{Q}_p(t_0)$ is the initial used flexibility value given by
\begin{equation}
    \Delta\dot{Q}_p(t_0) = C({T_B}_{nom}-T_B(t_0))
\end{equation}
In this paper, it is assumed that $\dot{Q}_{out}$ is a known profile due to the limited range of $\Delta T_L$ and $\Delta T_U$. Additionally, any errors in this profile can be compensated for by  updating $\Delta\dot{Q}_p(t_0)$ at each control step.
\section{Hierarchical Optimization of Demand}
\begin{figure}
    \centering
    \includegraphics[width = 1\columnwidth]{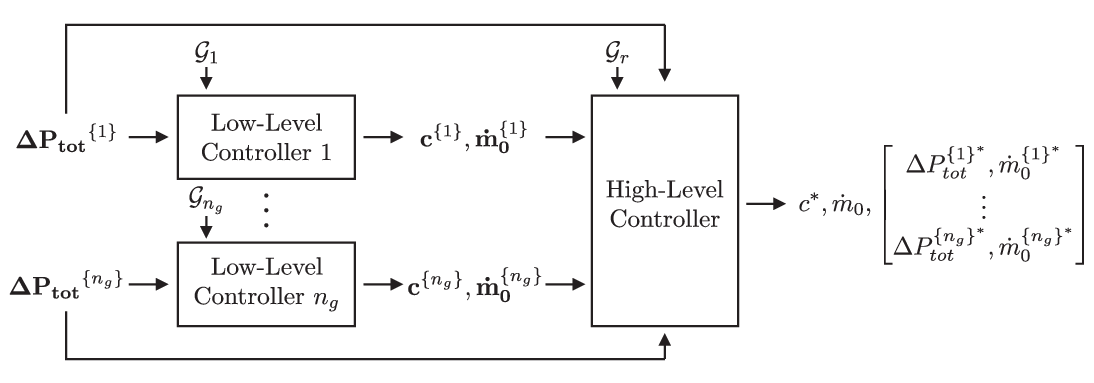}
    \caption{Proposed non-iterative hierarchical control structure.}
    \label{fig:diagram}
\end{figure}

To reduce communication overhead and ensure the problem is scalable, a novel hierarchical structure is proposed where the full network is decomposed into subsystems, each with a local controller. These local controllers serve to exploit the local building flexibility to optimize the subsystems behavior over a set of potential total pressure losses, $\Delta P_{tot}$ in each subsystem. By considering the flexibility at the local level, this hierarchical structure allows for improved performance within a subsystem while ensuring all other subsystems receive the flow needed to meet user demands via the high level controller. After the local controllers solve for the minimum cost behavior over a given set of $\Delta P_{tot}$ for each subnetwork, this information is passed to the high-level controller, which determines the best $\Delta P_{tot}$ value for each subsystem to minimize the total cost while maintaining a total pressure and mass flow balance in the network. A diagram of the proposed framework is shown in \cref{fig:diagram}.

\subsection{Partitioning of Network}
The first step to designing the low-level controllers is partitioning the network into $n_g$ smaller sub-networks. In this paper, the normalized cut metric is used to recursively bi-partition the overall network graph into subsystems. This method is commonly used in the partitioning of large-scale systems for distributed control \cite{jogwarDistributedControlArchitecture2019} and is easily integrated into the the graph-based modeling approach used for the DHN. The normalized cut metric  minimizes the cut cost as a fraction of the total edge weight \cite{shiNormalizedCutsImage2000}, given by
\begin{multline}
    Ncut(\mathcal{G}_i,\mathcal{G}_j) =\left(\sum_{i\in\mathcal{V}_i,j\in\mathcal{V}_j}w_{ij}\right)\times\\
    \left(\frac{1}{\sum_{i\in\mathcal{V}_i,k\in\mathcal{V}}w_{ik}}+\frac{1}{\sum_{j\in\mathcal{V}_j,k\in\mathcal{V}}w_{jk}}\right)
\end{multline}
where $\mathcal{G}_i = (\mathcal{V}_i,\mathcal{E}_i),\mathcal{G}_j= (\mathcal{V}_j,\mathcal{E}_j)$ are the two resulting subgraphs, and $w_{i,j}$ is the edge weight from node $i$ to $j$.\par
The optimal normalized cut can be approximated by solving the Rayleigh quotient minimization problem, which is solved using the spectral decomposition of the graph Laplacian via the Fiedler vector \cite{biyikoguLaplacianEigenvectorsGraphs2007}. The Fiedler vector is defined as the eigenvector associated with the second smallest eigenvalue of the graph Laplacian \cite{fiedlerAlgebraicConnectivityGraphs1973}. For a weighted, undirected graph, the graph Laplacian, $L$, can be found according to 
\begin{equation}
    L = I-(D^{-1})^{1/2}\Gamma_{ud}(D^{-1})^{1/2}
\end{equation}
where $I$ is an appropriately sized identity matrix, $\Gamma_{ud}$ is the weighted, undirected adjacency matrix, and $(D^{-1})^{1/2}$ is the square root of the inverse of $D$, where $D$ is the weighted degree matrix given by
\begin{equation}
    D_{ij} =\begin{cases}
    \sum_{k}w_{ik},\quad \forall\ k \text{ s.t. } e_{ik}\in\mathcal{E}& \text{if } i=j\\
    0 &\text{otherwise}
    \end{cases}
\end{equation}
For the decomposition of DHNs presented here, the adjacency matrix for the undirected graph is created from the original directed graph according to $\Gamma_{ud} = \Gamma^T\Gamma$, and the edge weights are chosen as $1/\rho V$, the constant term in $c_1$, the coupling coefficient for the temperature in the pipes, indicating the impact the in-flow temperature has on the pipe temperature.\par
The normalized cut metric relies on removing edges to create two disjoint subgraphs, with no overlapping nodes. However, because the edges represent pipes, removing these edges impacts the pressure losses and temperature dynamics of the resulting subsystems. Therefore, these neglected edges are added back into the appropriate subsystem, based on the root and terminal nodes of each subsystem, resulting in subgraphs with overlapping sets of nodes. \par
\subsection{Low-Level Optimization Problem}
The partitioned network can be represented by a set of graphs $\{\mathcal{G}^{\{1\}}\dots\mathcal{G}^{\{n_g\}}\}$, each with their own incidence matrix. For each each subsystem $j$, the following optimization problem is solved for a predetermined set of $\mathbf{\Delta P_{tot}}^{\{j\}} = [\Delta P_{tot\ 1}^{\{j\}}\dots \Delta P_{tot\ n_p(j)}^{\{j\}}]$.
\begin{subequations}
    \begin{equation}\label{eq:lm1} 
       c^{\{j\}}_i=\min_{\dot{m}_0^{\{j\}},\zeta_{U_j}>0} f(\dot{m}_{\mathcal{E}_j},T_{{\mathcal{E}_N}_j},\Delta\dot{Q}_p) \quad s.t. 
    \end{equation}
        
    \begin{equation}\label{eq:lm2}
        \Delta P_{\mathcal{E}_j} = \zeta_{\mathcal{E}_j} \dot{m}_{\mathcal{E}_j}^2,\quad \zeta_{U_j}\subset\zeta_{\mathcal{E}_j}
    \end{equation}
    \begin{equation}\label{eq:lm3}
        \Lambda \dot{m}_{\mathcal{E}_j} = \dot{m}_{\mathcal{V}_j}
    \end{equation}
    \begin{equation}\label{eq:lm4}
    \Delta P_{\mathcal{E}_j} = \Lambda^T P_{\mathcal{V}_j},\ P_{\mathcal{V}_j}(v_{root\ j})=\Delta P_{tot\ i}^{\{j\}}
    \end{equation}
    \begin{equation}\label{eq:lm5}
    \frac{d}{dt}T_{{\mathcal{E}_N}_j} = A(\dot{m}_{{\mathcal{E}_N}_j})T_{{\mathcal{E}_N}_j} + B \begin{bmatrix}{T_0}_j\\T_{setR}\\T_{amb}\end{bmatrix}
    \end{equation}
    \begin{equation}\label{eq:lm6}
    \dot{Q}_p = \dot{m}_{{\mathcal{E}_j}_U}c_p(T_{in\ {\mathcal{E}_j}_U}-T_{setR})
    \end{equation}
    \begin{equation}\label{eq:lm7}
    C\Delta T_L\leq\int_{t_0}^{t_f} \Delta\dot{Q}_p(t)\ dt+\Delta\dot{Q}_p(t_0)\leq C\Delta T_U
    \end{equation}
\end{subequations}
where \cref{eq:lm1} is the cost to be minimized. \Cref{eq:lm2,eq:lm3,eq:lm4} calculate the mass flow rates throughout the subgraph and \cref{eq:lm5,eq:lm6} calculates the heat delivered to the buildings in the partition, where ${T_0}_j$ is the local supply temperature and $T_{{in\ \mathcal{E}_j}_U}\subset T_{\mathcal{E}_j}$. Finally, \cref{eq:lm7} is a vector inequality that ensures the heat delivered to all buildings in the subsystem remain within their flexibility envelopes. The problem is optimized over the friction coefficients for the user edges, $\zeta_{U_j}$. Each user has a control valve that, when the position is modulated, changes the friction coefficient, thereby changing the mass flow rate to the building. This optimization is solved for the set $\Delta P_{tot}^{\{j\}}$ giving a mapping from  $\mathbf{\Delta P_{tot}}^{\{j\}}$ to the optimal costs ($\mathbf{c}^{\{j\}}$) and initial mass flow rate ($\mathbf{{\dot{m}_0}}^{\{j\}}$), where $\dot{m}_0\in\dot{m}_{\mathcal{E}_j}$, for each subsystem. This information is transmitted to the high-level controller to resolve the total network cost.
\subsection{High-Level Optimization Problem}
For the high-level controller, a reduced graph, $\mathcal{G}_r = (\mathcal{V}_r,\mathcal{E}_r)$,  is created, with a reduced incidence matrix $\Lambda_r$. The root and terminal node of each subgraph are represented by the nodes of the reduced graph, and the entire subsystem is reduced to a single edge with an associated set of pressure drops, costs, and initial mass flow rates triplets. From these triplets, the following integer-programming optimization problem is solved by the high-level controller:
\begin{subequations}
\begin{gather}
    \min_{\mathcal{I}=[i_1\dots i_{n_g}]} \sum_{j=1}^{n_g} c^{\{j\}}_{\mathcal{I}(j)} \quad s.t.\\
    \Lambda_r \begin{bmatrix}\dot{m}_{\mathcal{I}}\\ \dot{m}_{\mathcal{E}_r\cap e_{1\dots n_g}}\end{bmatrix} = \dot{m}_{\mathcal{V}_r}\label{eq:mdotred}\\
    \begin{bmatrix}\Delta P_{\mathcal{I}}\\ \Delta P_{\mathcal{E}_r\cap e_{1\dots n_g}}\end{bmatrix} - \Lambda_r^T P_{\mathcal{V}_r}< \epsilon\label{eq:pbalH}
\end{gather}
\end{subequations}
where $\mathcal{I}$ is a vector of the selected pressure loss indexes and the vectors in \cref{eq:mdotred} and \cref{eq:pbalH} are vectors of the chosen values for each subsystem, appended with edges $e_{1\dots n_g}$, which are subsystems that do not contain any users, and epsilon is a small constant, indicating an imperfect pressure balance, due to the discretization of potential pressure drops.
The result of this optimization problem is transmitted to the low-level controllers, and the optimal valve positions are implemented based on the desired friction coefficients. Additionally, the total initial mass flow rate, found from \cref{eq:mdotred} is transmitted to the heating plant. This optimization problem can then be solved in a receding horizon fashion to account for variations in heat demand, ambient temperature, and supply temperature from the forecasted conditions.
\section{Results}
\begin{figure}
    \centering
    \includegraphics[width = 1\columnwidth]{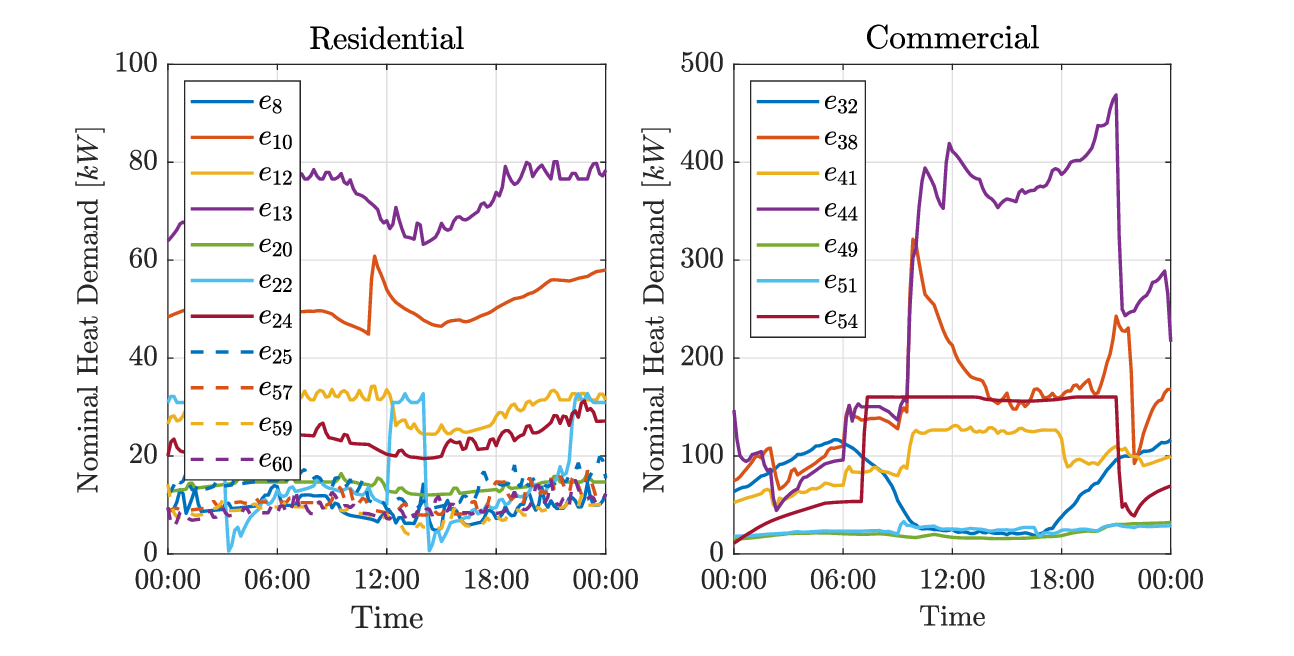}
    \caption{Nominal heat demand of residential and commercial users.}
    \label{fig:demand}
\end{figure}
\subsection{Description of Case Study}
The buildings' nominal heating demands (\cref{fig:demand}) for February 2nd and corresponding OpenStock models were obtained from NREL's ResStock \cite{horowitz2019resstock} and ComStock \cite{parker2023comstock} for a variety of buildings in Cook County, Illinois, presented in \cref{tbl:build}. OpenBuild \cite{gorecki2015openbuild} was used to develop state-space models of each building, from which the heat capacity was calculated using its step responses. The ambient temperature in typical meteorological year 3 (TMY3) was provided by NREL in the 2021 data set\cite{horowitz2019resstock}.\par
The DHN graph was created using a variety of typical network characteristics, tabulated in \cref{tbl:network} and an illustrative DHN layout, shown in \cref{fig:graph} to contain a variety of network configurations seen in real-world DHNs. The network supply temperature, $T_0$ was assumed to be constant. The simulation was performed using the parallel computing toolbox in the Matlab environment to solve the low-level control problem simultaneously using CasADi \cite{Andersson2019} and IPOPT on a system equipped with an Intel Core i7-8565U CPU.
\begin{table}
\caption{Parameters for network.}
\label{tbl:network}
\centering
\begin{tabular}{l c c c}
\toprule
Parameter & Symbol & Value & Units\\
\midrule
Density &$\rho$ & 971 & $kg/m^3$\\
Specific heat capacity & $c_p$ & 4179 & $J/kgK$\\
Supply temperature & $T_0$ & 80 & $C$\\
Ambient temperature & $T_{amb}$ & -19.5 -- -13.9 & $C$\\
Pipe diameter & $D$ & 0.15 -- 0.40 & $m$\\
Pipe length & $L$ & 10 -- 100 & $m$\\
Bypass length & $L$ & 3 & $m$\\
Heat transfer coefficient & $h$ & 1.5 & $W/m^2K$\\
Friction coefficient & $\lambda$ & 0.01 & -\\
\bottomrule
\end{tabular}
\end{table}

\begin{table}
\caption{Parameters for buildings.}
\label{tbl:build}
\centering
\begin{tabular}{c c l c c}
\toprule
$\mathcal{E}$ & Building ID & Type & Area ($m^2$) & $C$ ($MJ/K$)\\
\midrule
$e_8$ & R-3561 & House & 160 & 78\\
$e_{10}$ & R-80372 & House & 760 & 400\\
$e_{12}$ & R-3801 & House & 160 & 900\\
$e_{13}$ & R-80387 & House & 250 & 326\\
$e_{20}$ & R-80368 & House & 110 & 526\\
$e_{22}$ & R-4017 & House & 760 & 251\\
$e_{24}$ & R-4090 & House & 160 & 464\\
$e_{25}$ & R-4177 & House & 250 & 818\\
$e_{32}$ & C-177428 & Medical & 7000 & 12562\\
$e_{38}$ & C-1700 & Retail & 3500 & 9871\\
$e_{41}$ & C-232839 & Retail & 1600 & 6059\\
$e_{44}$ & C-343832 & Retail & 3500 & 8575\\
$e_{49}$ & C-18740 & Warehouse & 1600 & 2393\\
$e_{51}$ & C-123604 & Office & 700 & 1511\\
$e_{54}$ & C-95364 & Retail & 7000 & 5088\\
$e_{57}$ & R-20041 & Apartment & 80 & 513\\
$e_{59}$ & R-28770 & Apartment & 110 & 265\\
$e_{60}$ & R-22719 & Apartment & 110 & 657\\
\bottomrule
\end{tabular}
\end{table}

\begin{figure}
    \centering
    \includegraphics[width = 1\columnwidth]{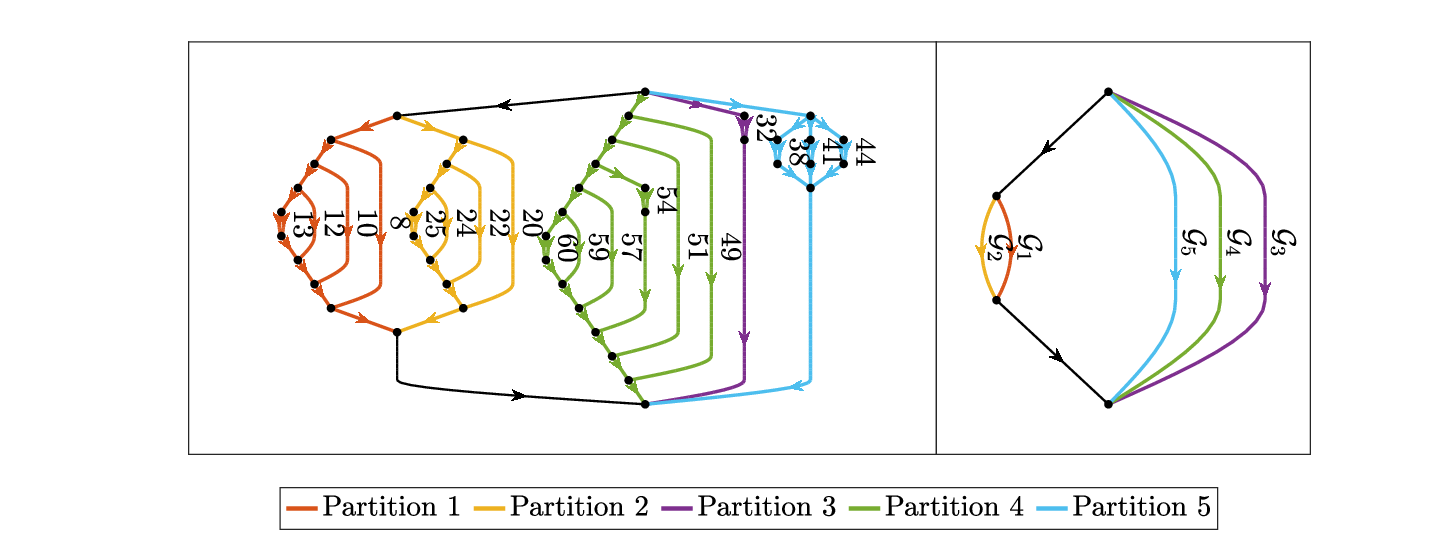}
    \caption{Partitioned network graph and resulting reduced graph.}
    \label{fig:part}
\end{figure}

The network, partitioned into five sub-networks with the resulting reduced graph to be considered by the high-level controller is shown in \cref{fig:part}. These partitions correspond with the major branches of the network and the cuts are made through the longest pipe segments, as longer edges decrease the coupling between neighboring nodes. Due to the chosen network configuration, the low-level controller assumes the local supply temperature is equal to be the global supply temperature, but this assumption can be modified based on network data.\par
The cost function for the subsystems in \cref{eq:lm1} was chosen to be
\begin{equation}
    f_j(\dot{m}_{\mathcal{E}_j},T_{{\mathcal{E}_N}_j},\Delta\dot{Q}_p)=\int_{t=0}^{t_f} \sum \dot{m}_{{\mathcal{E}_j}_{By}}\ dt
\end{equation}
where $\mathcal{E}_{j_{By}}\subset \mathcal{E}_j$ are the bypass edges, which corresponds to minimizing the flow that is directly recirculated to the return network, but this cost can be modified to consider pumping costs, heat losses or any other desired network metric. The total pressure losses considered for an individual subsystem ranged between 0.5 Pa and 300 Pa. The 24 hour simulation was implemented in a receding horizon fashion where the control horizon one hour with a sampling rate of 10 minutes. Over this one hour optimization horizon, the pressure loss was held constant for the first control interval of 10 minutes, and was unconstrained during the remaining time. The removal of the pressure loss constraint after the first control interval allowed for the optimization to consider upcoming demands, without over-constraining the feasible solution set, as the solution beyond the current control step does not have to be consistent throughout the network. The resulting initial mass flow rate and $\zeta_U$ values for the control interval were then used in a simulation for the total network response every ten minutes. The simulation accounts for the true pressure balance in the network, eliminating the issues that could be caused by the imperfect pressure balance seen in \cref{eq:pbalH} and ensuring the flexibility envelopes will not be exceeding during real operation due to this imperfect balance.\par
\subsection{Results}
\begin{figure}
    \centering
    \includegraphics[width = .8\columnwidth]{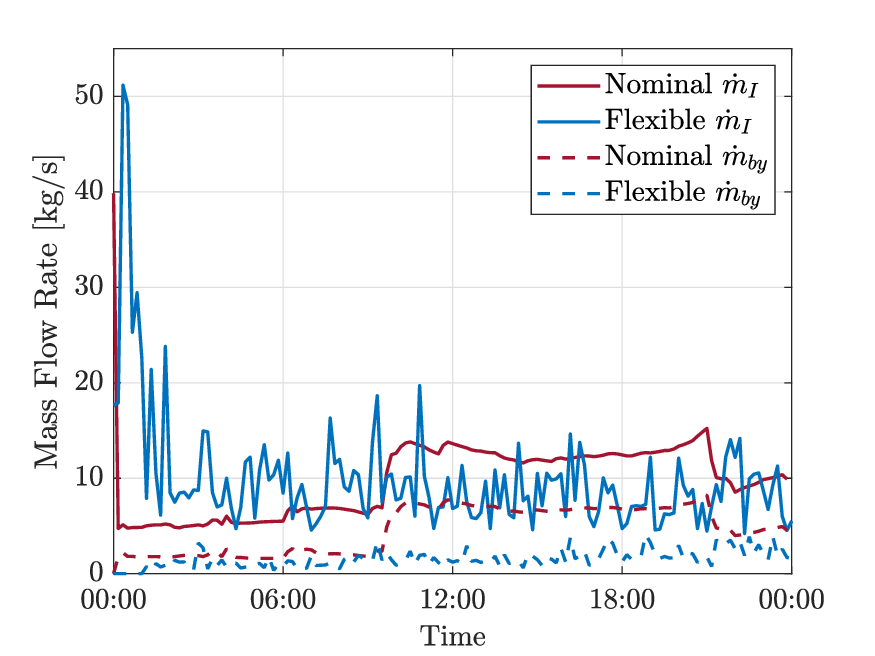}
    \caption{Supplied mass flow and bypass mass flow in the nominal and optimized case.}
    \label{fig:mI}
\end{figure}

\begin{figure}
    \centering
    \includegraphics[width = 1\columnwidth]{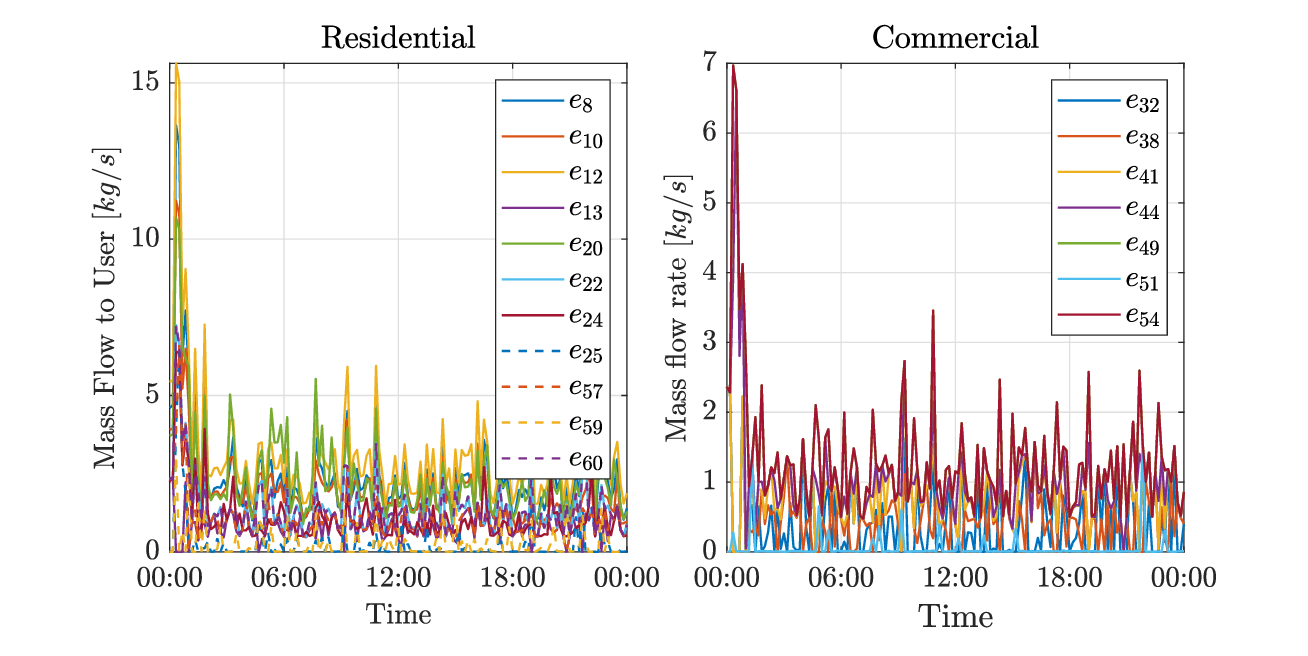}
    \caption{Mass flow rate provided to residential and commercial users.}
    \label{fig:mdot}
\end{figure}

\begin{figure}
    \centering
    \includegraphics[width = 1\columnwidth]{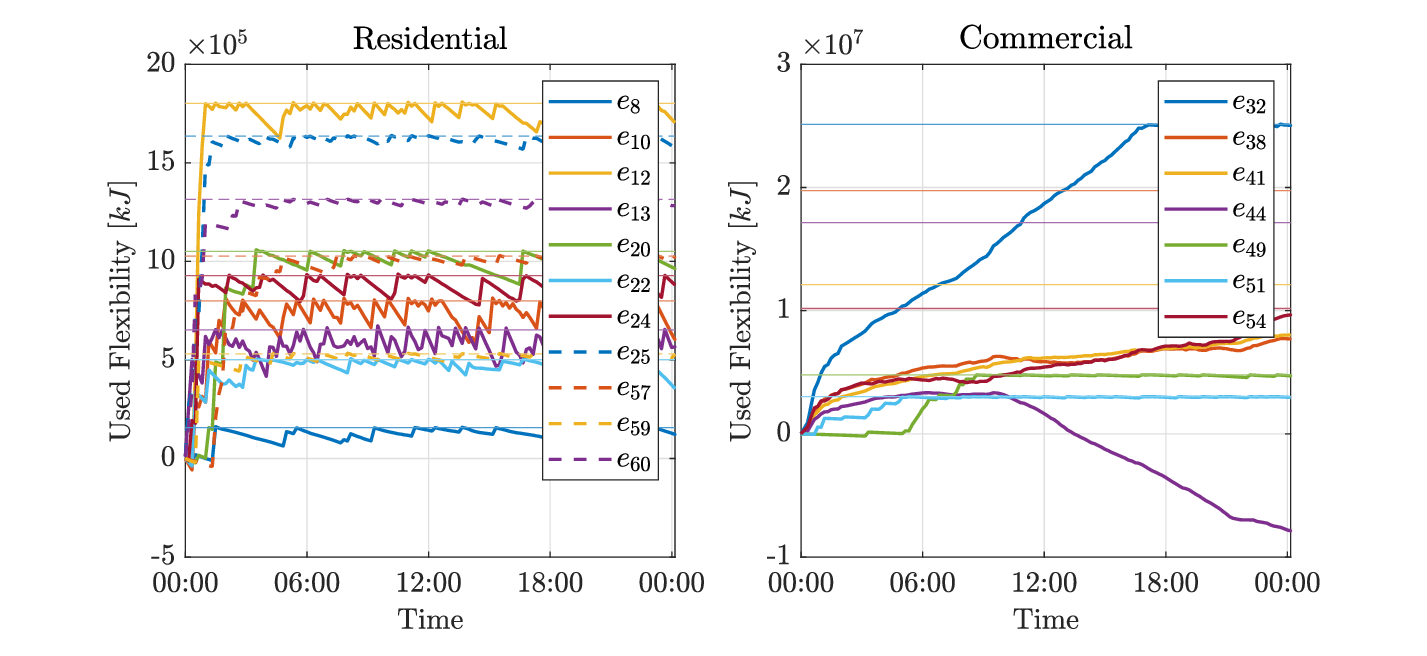}
    \caption{Used flexibility of residential and commercial users with limits labeled.}
    \label{fig:flex}
\end{figure}

\begin{figure}
    \centering
    \includegraphics[width = 1\columnwidth]{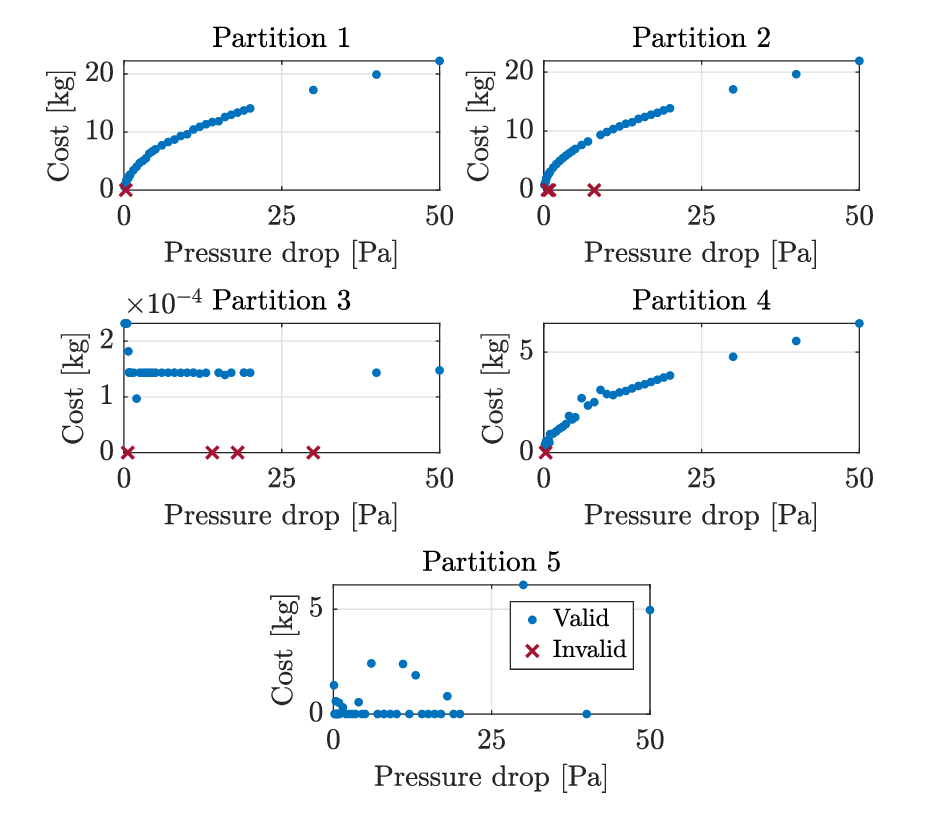}
    \caption{Low level costs for considered pressure drops in all subsystems at optimization step 72.}
    \label{fig:costs}
\end{figure}

In supplying the users with their nominal heat demand, where there was no flexibility in the heat supplied, the integral of the sum of the bypass mass flow rates was $4.01\times 10^5$ kg. In the optimized case, this cost was $1.33\times 10^5$ kg, a 67\% reduction in total cost. The costs as a function of time can be seen in \cref{fig:mI}. It is also noted that, while the bypass mass flow rate decreased, the total mass supplied to the network increased by 2.2\% from $8.48\times 10^5$ kg in the nominal case to $8.66\times 10^5$ kg in the optimized case. The resulting mass flow rates for all the buildings in the DHN are shown in \cref{fig:mdot}. The used flexibility is shown in \cref{fig:flex}, where the horizontal lines represent the upper limits of the flexibility envelopes. The cost for the low level subsystems for each pressure loss value at optimization step 72 can be seen in \cref{fig:costs}. Here, invalid results occurred if the problem constraints can not be met while maintaining the desired pressure drop. The average simulation time for a single partition for the one hour optimization horizon was 7.4 minutes, allowing the presented optimization scheme to be successfully implemented in real time.
\section{Conclusions}
This paper provides a framework for implementing a hierarchical optimization scheme for the demand-side control of large-scale district heating networks. It relies on the flexibility provided by individual network consumers to minimize the subsystems' costs while ensuring the building temperatures remain in an acceptable range. The low-level controllers optimize this cost given a set pressure losses and find the mass flow rate provided to the subsystem. The high-level controller then selects the optimal mass flow, pressure pairs to optimize the entire network's performance. The hierarchical framework was demonstrated in the open-loop optimization of a medium-scale 18 user DHN, with the goal of minimizing bypass flow. The results were compared to the case where the nominal heat demand was provided to each user and resulted in an 67\% reduction in unused mass flow. Future work will be to expand the hierarchical framework to consider more complex network configurations with additional heat sources and pumps. Additionally, with the selected control objective, the total mass flow needed by the network increased due to the tendency for the optimization problem to heat the buildings to the upper limit of their flexibility envelope. Future work will look to add a penalty term to the cost function to prevent this undesired behavior and the associated initial mass flow spike. It will also look to consider additional cost elements such as pumping costs, heat losses, and the total heat supplied to the network.


\bibliographystyle{IEEEtran}
\bibliography{sources}

\end{document}